\RequirePackage{fix-cm}
\documentclass[twocolumn,epjc3]{svjour3}
\smartqed  
\RequirePackage{graphicx}
 \RequirePackage{mathptmx}      
\RequirePackage[numbers,sort&compress]{natbib}

\usepackage{amsmath}
\usepackage{amssymb}

\begin{document}

\title{Improved upper bounds on Kaluza-Klein gravity with current Solar System experiments and observations
}


\author{Xue-Mei Deng\thanksref{addr1,e1}
        \and
        Yi Xie\thanksref{addr2,addr3,e2} 
}

\thankstext{e1}{e-mail: xmd@pmo.ac.cn}
\thankstext{e2}{e-mail: yixie@nju.edu.cn}


\institute{Purple Mountain Observatory, Chinese Academy of Sciences, Nanjing 210008, China \label{addr1}
           \and
           School of Astronomy and Space Science, Nanjing University, Nanjing 210093, China \label{addr2}
           \and
           Shanghai Key Laboratory of Space Navigation and Position Techniques, Shanghai 200030, China \label{addr3}
}

\date{Received: date / Accepted: date}

\maketitle

\begin{abstract}
	As an extension of previous works on classical tests of Kaluza-Klein (KK) gravity and as an attempt to find more stringent constraints on this theory, its effects on physical experiments and astronomical observations conducted in the Solar System are studied. We investigate the gravitational time delay at inferior conjunction caused by KK gravity, and use new Solar System ephemerides and the observation of \textit{Cassini} to strengthen constraints on KK gravity by up to two orders of magnitude. These improved upper bounds mean that the fifth-dimensional space in the soliton case is  a very flat extra dimension in the Solar System, even in the vicinity of the Sun.
\end{abstract}

\section{Introduction}

Gravitation was the first known fundamental force in the universe. However, gravitation still can not be included into a quantum framework such as the standard model of the strong, weak and electromagnetic interactions. It is an undoubtedly grand challenge to unify gravitation with the three others. Inspired by this issue, as candidates of unification theories, some gravitational theories with spacetime more than 4 dimensions try to bridge the gap between gravitation and other fundamental forces \cite[e.g][]{Blagojevic2001Book,Freedman2012Book}.

Among them, the five-dimensional  Kaluza-Klein (KK) gravity, originally proposed by Kaluza \cite{Kaluza1921SPAW.966} and Klein \cite{Klein1926ZPhy37.895}, intends to unify gravitation and electromagnetism by inducing one extra dimension in addition to the usual four-dimensional spacetime \cite[see][for reviews and references therein]{Duff1986PhyRep130.1,Overduin1997PhyRep283.303O,Appelquist1987Book,Wesson2006Book,Wesson2007Book}. Given the promising prospect, it is desirable to test KK gravity with physical experiments and astronomical observations \cite[e.g.][]{Gross1983NuPhB226.29,Davidson1985PLB155.247,Lim1995JMP36.6907,Liu2000ApJ538.386,Overduin2000PRD62.102001,Overduin2013GRG45.1723}. Here, a critical point is that, in order to describe the spacetime of the Solar System where these tests are conducted, there exist different approaches and classes of solutions in KK gravity, such as soliton case \cite{Gross1983NuPhB226.29,Sorkin1983PRL51.87,Davidson1985PLB155.247,Wesson1994CQG11.1341,Lim1995JMP36.6907,Kalligas1995ApJ439.548,Liu2000ApJ538.386} and Schwarzschild-like solution \cite{Myers1986AnPhy172.304,Rahaman2009IJTP48.3124}. The results obtained by different approaches might not agree with each other, which is caused by the freedom in choosing higher dimensional solutions to represent the Solar System in four dimensions \cite{Xu2007PLB656.165}. In this investigation, following the works of Refs. \cite{Lim1995JMP36.6907,Kalligas1995ApJ439.548,Liu2000ApJ538.386}, we focus on the soliton case of KK gravity. In the induced-matter picture \cite{Wesson1992JMP33.3883,Liu1992JMP33.3888,Wesson1994CQG11.1341} that matter and energy are induced in four-dimensional by pure geometry in five-dimensional spacetime, the soliton can have an extended matter distribution as a dark-matter candidate \cite{Wesson1994ApJ420.L49,Overduin2004PhyRep402.267O}.

In KK gravity, the soliton metric \cite{Gross1983NuPhB226.29,Sorkin1983PRL51.87,Davidson1985PLB155.247} satisfies the five-dimensional vacuum field equations. It can reduce to the standard four-dimensional Schwarzschild solution on hypersurfaces with one extra dimensional coordinate being constant, and the metric has no explicit dependence of the extra dimension. Solitons are five-dimensional objects whose metric is static and spherically symmetric in ordinary space and asymptotically flat. The soliton metric has been generalized in a variety of ways \cite[see][for reviews]{Overduin1997PhyRep283.303O,Wesson2011arXiv1104.3244}. The correction of KK gravity to Einstein's general relativity (GR) in four-dimensional spacetime can be characterized by a parameter $b$, which needs to be determined by physical experiments and astronomical observations. Taking the soliton metric and choosing $b$ as an independent parameter, the authors of Ref. \cite{Liu2000ApJ538.386} calculated the leading contributions of perihelion shift of Mercury, light deflection, gravitational time delay at superior conjunction (SC) and geodetic precession caused by KK gravity. They found that $|b|<0.07$ in the Solar System based on the results of corresponding measurements \cite{Liu2000ApJ538.386}. More recently, by using measurements of geodetic precession from Gravity Probe B \cite{Everitt2011PRL106.221101}, the authors of Ref. \cite{Overduin2013GRG45.1723} obtained a new upper limit as $|b|<0.02$.

In this work, we improve and extend these previous works in the following prospectives. First, we investigate gravitational time delay at inferior conjunction (IC) caused by KK gravity, which was not considered in the previous works. The authors of Ref. \cite{Liu2000ApJ538.386} calculated the time delay at SC and found their best upper bound as $|b|<0.07$. However, the time delay measurements are highly dominated by the noise due to solar corona and it is extremely difficult to disentangle this noise from others. The situation at IC is totally different. The biggest uncertainties come from the positions of the receiver and the emitter. With this advantage, it was adopted to constrain the $f(T)$ gravity \cite{Xie2013MNRAS433.3584}. But we find that the time delay at IC can not yield an useful bound on $b$.

Second, the \textit{Cassini} superior conjunction (SC) experiment \cite{Bertotti2003Nature425.374} is, for the first time, used to test KK gravity. Around the time of a solar conjunction in 2002, the experiment was carried out to measure the fractional frequency shift for a two-way radio signal. In order to overcome the solar corona noise, \textit{Cassini} used a multi-frequency link in which three different phases were measured at the ground station \cite{Bertotti1993AA269.608,Bertotti1998SolPhy178.85}. This experiment was also re-modeled to test some modified theories of gravity \cite[e.g.][]{Kagramanova2006PLB634.465,Xie2013MNRAS433.3584}. We calculate the fractional frequency shift caused by KK gravity and obtain a new upper bound on $b$, which is tighter than the one of Ref. \cite{Liu2000ApJ538.386} by about 10 times.

Third, we will improve the upper bound on $b$ in the Solar System by making use of current and highly accurate datasets of the planetary motions. And we will also try to reduce the contamination in our investigation due to the uncertainty of the Sun's quadrupole moment, which affects the motion of Mercury significantly \cite{Nobili1986Nature320.39}. For these purposes, we will use the supplementary advances of the perihelia provided by INPOP10a (IMCCE, France) \cite{Fienga2011CMDA111.363} and EPM2011 (IAA RAS, Russia) \cite{Pitjeva2013SSR47.386} ephemerides. These two ephemerides were recently adopted in planetary science \cite{Iorio2012CMDA112.117,Iorio2014MNRAS444.L78} and in detecting gravitational effects and testing modified theories of gravity \cite{Iorio2012MNRAS427.1555,Iorio2013CQG30.165018,Xie2013MNRAS433.3584,Iorio2014IJMPD23.1450006,Iorio2014MNRAS437.3482,Iorio2014CQG31.085003,Deng2014ApSS350.103,Deng2015NewAst35.36,Deng2015IJTP54.1739,Deng2015AnnPhy361.62,Zhao2015PRD92.064033}. In order to find a clearer bound, we will also take the Lense-Thirring effect due to the Sun's angular momentum into account. Neither of two factors are considered in the previous work of Ref. \cite{Liu2000ApJ538.386}, although the effect of the Sun's quadrupole moment was included. With these efforts, the upper bounds on $b$ we obtain are improved by at least 2 orders of magnitude with respect to the one of Ref. \cite{Liu2000ApJ538.386}.

The remainder of the paper is organized as follows. Section \ref{sec:model} is devoted to describing the effects of KK gravity on physical experiments and astronomical observations. In Section \ref{sec:bounds}, we confront these effects with available datasets and estimate their upper bounds on $b$. Finally, in Section \ref{sec:con}, we conclude and discuss our results.

\section{Effects of KK gravity on experiments and observations}

\label{sec:model}

Some effects of KK gravity on physical experiments and astronomical observations were well modeled in Ref. \cite{Liu2000ApJ538.386}, including perihelion shift, light deflection and gravitational time delay at SC. In this section, we will first briefly review these classical tests of KK gravity for completeness \cite[see][for details]{Liu2000ApJ538.386}. And, then, the gravitational time delay at IC and the \textit{Cassini} SC experiment will be modeled.

\subsection{Classical tests}

\subsubsection{Perihelion shift}

KK gravity can induce an additional perihelion shift of a planet around the Sun, which is \cite{Liu2000ApJ538.386}
\begin{equation}
  \dot{\omega}_{\,\mathrm{KK}} = 3\frac{GM_{\odot}n_{\mathrm{P}}}{c^{2}a_{\mathrm{P}}(1-e_{\mathrm{P}}^{2})}\bigg(\frac{b}{6}+\sqrt{1-\frac{3b^{2}}{4}}\bigg),
\end{equation}
where $M_{\odot}$ is the mass of the Sun, $a_{\mathrm{P}}$ is the semi-major axis of the planet, $e_{\mathrm{P}}$ is its eccentricity and $n_{\mathrm{P}}$ is its Keplerian mean motion. The observed deviation from GR caused by KK gravity in $\dot{\omega}$ as
\begin{eqnarray}
  \label{dPS}
  \delta^{\mathrm{GR}}_{\dot{\omega}} & \equiv & |\dot{\omega}_{\,\mathrm{KK}}- \dot{\omega}_{\,\mathrm{GR}}|\nonumber\\
  & = & 3\frac{GM_{\odot}n_{\mathrm{P}}}{c^{2}a_{\mathrm{P}}(1-e_{\mathrm{P}}^{2})}\bigg|\frac{b}{6}+\sqrt{1-\frac{3b^{2}}{4}}-1\bigg|.
\end{eqnarray}

\subsubsection{Light deflection}

Besides the part caused by GR, the contribution in light delfection from the fifth-dimensional space in KK gravity can be obtained as \cite{Liu2000ApJ538.386}
\begin{equation}
  \Delta\phi_{\,\mathrm{KK}} = 4\frac{GM_{\odot}}{c^{2}d}\sqrt{1-\frac{3b^{2}}{4}},
\end{equation}
where $d$ is the closest approach of the light ray. The deviation in the light deflection from GR is
\begin{equation}
\label{dLD}
\delta^{\mathrm{GR}}_{\Delta\phi} \equiv |\Delta\phi_{\,\mathrm{KK}}-\Delta\phi_{\,\mathrm{GR}}| = 4\frac{GM_{\odot}}{c^{2}d}\bigg|\sqrt{1-\frac{3b^{2}}{4}}-1\bigg|.
\end{equation}

\subsubsection{Time delay at SC}

In the case of SC, when the receiver is on the opposite side of the Sun as seen from the emitter, the round-trip time delay in KK gravity is \cite{Liu2000ApJ538.386}
\begin{equation}
	\Delta t^{\mathrm{KK}}_{\mathrm{SC}} = \frac{2}{c}(r_{\mathrm{E}}+r_{\mathrm{R}})+4\frac{GM_{\odot}}{c^{3}}\sqrt{1-\frac{3b^{2}}{4}}\ln\bigg(\frac{4r_{\mathrm{E}}r_{\mathrm{R}}}{d^{2}}\bigg),
\end{equation}
where $r_{\mathrm{E}}$ is the distance between the emitter and the Sun and $r_{\mathrm{R}}$ is the distance between the reflector and the Sun. The deviation caused by KK gravity is
\begin{eqnarray}
\label{dTDSC}
  \delta^{\mathrm{GR}}_{\Delta t_{\mathrm{SC}}} & \equiv & |\Delta t^{\mathrm{KK}}_{\mathrm{SC}}-\Delta t^{\mathrm{GR}}_{\mathrm{SC}}|\nonumber\\
 & = & 4\frac{GM_{\odot}}{c^{3}}\bigg|\sqrt{1-\frac{3b^{2}}{4}}-1\bigg|\ln\bigg(\frac{4r_{\mathrm{E}}r_{\mathrm{R}}}{d^{2}}\bigg).
\end{eqnarray}

\subsection{Time delay at IC}

The time delay at SC are highly dominated by the noise of solar corona, which is extremely difficult to disentangle from others. So any constraint obtained at that period with ranging measurements has to be taken with a lot of caution and a proper test should be done together with fitting the solar corona model used for the analysis of these data during the SC period \cite[see][for details]{Verma2013AA550.A124}.

However, the situation at IC is totally different and it was \emph{not} considered in the previous work of Ref. \cite{Liu2000ApJ538.386}. At IC, the reflector, which is usually a spacecraft with a radio transponder, is between the emitter and the Sun. The biggest uncertainties come from the positions of the receiver and the emitter, which range from a few centimeter to several hundreds meters. Following similar procedure like the one of Ref. \cite{Liu2000ApJ538.386}, we can obtain the round-trip time delay at IC as
\begin{eqnarray}
	\label{tt0}
  \Delta t^{\mathrm{KK}}_{\mathrm{IC}} & = & \frac{2}{c}\sqrt{r^{2}_{\mathrm{E}}-d^{2}}-\frac{2}{c}\sqrt{r^{2}_{\mathrm{R}}-d^{2}}\nonumber\\
  & & +4\frac{GM_{\odot}}{c^{3}}\bigg(a+\frac{b}{2}\bigg)\ln\bigg(\frac{r_{\mathrm{E}}+\sqrt{r^{2}_{\mathrm{E}}-d^{2}}}{d}\bigg)\nonumber\\
  & & -4\frac{GM_{\odot}}{c^{3}}\bigg(a+\frac{b}{2}\bigg)\ln\bigg(\frac{r_{\mathrm{R}}+\sqrt{r^{2}_{\mathrm{R}}-d^{2}}}{d}\bigg).
\end{eqnarray}
By making use of the conditions $r_{\mathrm{E}}\gg d$ and $r_{\mathrm{R}}\gg d$ again, we can have
\begin{equation}
\label{tt}
\Delta t^{\mathrm{KK}}_{\mathrm{IC}} = \frac{2}{c}(r_{\mathrm{E}}-r_{\mathrm{R}})+4\frac{GM_{\odot}}{c^{3}}\sqrt{1-\frac{3b^{2}}{4}}\ln\bigg(\frac{r_{\mathrm{E}}}{r_{\mathrm{R}}}\bigg),
\end{equation}
where $d$ is cancelled out because of the minus signs in Eq. (\ref{tt0}). When $b=0$, Eq. (\ref{tt}) matches the one of GR shown in Ref. \cite{Nelson2011Metrologia48.171}. We can also obtain
\begin{eqnarray}
\label{dTDIC}
  \delta^{\mathrm{GR}}_{\Delta t_{\mathrm{IC}}} & \equiv & |\Delta t^{\mathrm{KK}}_{\mathrm{IC}}-\Delta t^{\mathrm{GR}}_{\mathrm{IC}}|\nonumber\\
  & = & 4\frac{GM_{\odot}}{c^{3}}\bigg|\sqrt{1-\frac{3b^{2}}{4}}-1\bigg|\ln\bigg(\frac{r_{\mathrm{E}}}{r_{\mathrm{R}}}\bigg).
\end{eqnarray}
This equation will be used in the Sec. \ref{sec:bounds} to estimate a new upper bound on $b$.

\subsection{\textit{Cassini} SC experiment}

In this subsection, we will, for the first time, take the \textit{Cassini} SC experiment \cite{Bertotti2003Nature425.374} to test KK gravity. Between 6 June to 7 July 2002 around the time of a solar conjunction, when \textit{Cassini} was on its way to Saturn, the experiment was carried out. In the experiment, what is measured is not the time delay but the relative change in the frequency. Around the SC moment, a ground station transmitted a radio-wave signal with frequency $\nu_{0}$ to the spacecraft. This signal was coherently transponded by the spacecraft and sent back to the Earth. In order to overcome the solar corona noise, \textit{Cassini} used high-frequency carrier waves in the Ka-band, in addition to the X-band for standard operation, and a multi-frequency link in which three different phases were measured at the ground station \cite{Bertotti1993AA269.608,Bertotti1998SolPhy178.85}. It was found \cite{Bertotti2003Nature425.374} that $\gamma-1=(2.1\pm2.3)\times10^{-5}$, where the parameter $\gamma$ is a parametrized post-Newtonian (PPN) parameter and it determines how much space curvature produced by unit rest mass \cite{Will1993TEGP,Will2006LRR9.3}.

For testing KK gravity, we have to re-model the two-way fractional frequency fluctuation by taking the influence from the fifth dimension. The fractional frequency shift at SC is \cite{Iess1999CQG16.1487}
\begin{eqnarray}
  y_{\mathrm{SC}}\equiv\frac{\nu(t)-\nu_{0}}{\nu_{0}}=\frac{\mathrm{d}\Delta t_{\mathrm{SC}}}{\mathrm{d}t},
\end{eqnarray}
where the contribution owing to the KK gravity is
\begin{eqnarray}
  y^{\mathrm{KK}}_{\mathrm{SC}}&=&\frac{\mathrm{d}\Delta t^{\mathrm{KK}}_{\mathrm{SC}}}{\mathrm{d}t} =-8\frac{GM_{\odot}}{c^{3}d}\sqrt{1-\frac{3b^{2}}{4}}\frac{\mathrm{d}d(t)}{\mathrm{d}t}.
\end{eqnarray}
In the \textit{Cassini} SC experiment, $\mathrm{d}d(t)/\mathrm{d}t$ was close to the orbital velocity of the Earth, $v_{\oplus}$. The experiment started 12 days before the SC and ended 12 days after it. In one day, the distance of closest approach of the signal changes by about $1.5R_{\odot}$, where $R_{\odot}$ denotes the radius of the Sun. Thus for gravitational frequency shift, the possible deviation from GR for KK gravity in this experiment is
\begin{eqnarray}
\label{dySC}
  \delta^{\mathrm{GR}}_{y_{\mathrm{SC}}} & \equiv & |y^{\mathrm{KK}}_{\mathrm{SC}}-y^{\mathrm{GR}}_{\mathrm{SC}}|\nonumber\\
  & = & |y^{\mathrm{KK}}_{\mathrm{SC}}(12\mathrm{d})-y^{\mathrm{GR}}_{\mathrm{SC}}(12\mathrm{d})-y^{\mathrm{KK}}_{\mathrm{SC}}(0)+y^{\mathrm{GR}}_{\mathrm{SC}}(0)|\nonumber\\
&=&\frac{128GM_{\odot}}{27c^{3}R_{\odot}}\bigg|\sqrt{1-\frac{3b^{2}}{4}}-1\bigg|v_{\oplus}.
\end{eqnarray}
When $b=0$, this deviation caused by KK gravity vanishes. The above equation will be taken in the Sec. \ref{sec:bounds} to estimate a new upper bound on $b$.

\section{Improved and new upper bounds on $b$}

\label{sec:bounds}

\subsection{Perihelion shift}

In the Solar System, $\delta^{\mathrm{GR}}_{\left<\dot{\omega}\right>}$ of inner planets range from several tens to hundreds micro-arcseconds per century ($\mu$as cy$^{-1}$) \citep{Nordtvedt2000PRD61.122001,Pitjeva2005AstL31.340,Fienga2011CMDA111.363}. Based on Eq. (\ref{dPS}), we can obtain a bound as
\begin{eqnarray}
  \label{qdGRomega}
    -\frac{\delta^{\mathrm{GR}}_{\dot{\omega}}}{10\; \mu\mathrm{as}\; \mathrm{cy}^{-1}} & \le &
    6.39\times10^4 \bigg(\frac{a_{\mathrm{P}}}{\mathrm{au}}\bigg)^{-5/2}(1-e_{\mathrm{P}}^2)^{-1}\nonumber\\
     &&\times\bigg|b+6\sqrt{1-\frac{3b^{2}}{4}}-6\bigg| \nonumber\\
     & \le & \frac{\delta^{\mathrm{GR}}_{\dot{\omega}}}{10\; \mu\mathrm{as}\; \mathrm{cy}^{-1}},
\end{eqnarray}
where au is the astronomical unit.

In the case of the Solar System's planets, $\delta^{\mathrm{GR}}_{\dot{\omega}}$ is closely connected with the supplementary advances of the perihelia $\dot{\omega}_{\mathrm{sup}}$ provided by modern ephemerides, such as INPOP10a \cite{Fienga2010IAUS261.159,Fienga2011CMDA111.363} and EPM2011 \cite{Pitjeva2013SSR47.386,Pitjeva2013MNRAS432.3431,Pitjev2013AstL39.141}. INPOP10a and EPM2011 were obtained by fitting the ``standard model'' of dynamics to observational data, where ``standard model'' means the Newton's law of gravity and the Einstein's GR (apart from the Lense-Thirring effect of GR, see below for details). In INPOP10a and EPM2011 ephemerides, the ``standard model'' fitted to observations include not only dynamics of natural bodies and artificial spacecrafts, but also propagation of electromagnetic waves and how instruments onboard the spacecrafts and on Earth work. Therefore, KK gravity was modeled neither in INPOP10a nor in EPM2011, and the parameter $b$ was not determined in these least-square fittings.

These $\dot{\omega}_{\mathrm{sup}}$ might represent possibly mismodeled or unmodeled parts of perihelion advances according to the Newton's law and GR. They are almost all compatible with zero so that they can be used to draw bounds on quantities parametrizing unmodeled ``forces'' like KK gravity in this case. Nonetheless, the latest results by EPM2011 \cite{Pitjeva2013MNRAS432.3431,Pitjev2013AstL39.141} returned non-zero  values for Venus and Jupiter. Although the level of their statistical significance was not too high and further investigations are required, we still take them into account in this work. In the recent past, an extra non-zero effect on Saturn's perihelion was studied \cite{Iorio2009AJ137.3615}. And, the ratios of the non-zero values of the supplementary precessions of Venus and Jupiter by EPM2011 \cite{Pitjeva2013MNRAS432.3431,Pitjev2013AstL39.141} have been recently used to test a potential deviation from GR \cite{Iorio2014MNRAS437.3482}.

In the construction of $\dot{\omega}_{\mathrm{sup}}$ \cite[see][for details]{Fienga2010IAUS261.159}, the effects caused by the Sun's quadrupole mass moment $J^{\odot}_2$ are considered and isolated in the final results, but the perihelion shifts caused by the Lense-Thirring effect \cite{Lense1918PhyZ19.156} due to the Sun's angular momentum $S_{\odot}$ and caused by the uncertainty of the Sun's quadrupole moment are absent. In order to obtain a cleaner bound, we will not use the inequality (\ref{qdGRomega}) but the following equation that
\begin{equation}
  \label{domegasupp}
  \dot{\omega}_{\mathrm{sup}} =  \delta^{\mathrm{GR}}_{\dot{\omega}} + \dot{\omega}_{\,\mathrm{LT}} + \dot{\omega}_{\mathcal{J}_{\odot}}.
\end{equation}
It is worth mentioning that GR predicts the Sun can induce two kinds of perihelion shifts for a planet. Based on the analogy between gravitation and electromagnetism, the bigger one is called gravitoelectric \cite{Mashhoon2003GRQC0311030}, depending only on $M_{\odot}$; the smaller one is called gravitomagnetic \cite{Mashhoon2003GRQC0311030}, depending on $S_{\odot}$, which is also called the Lense-Thirring effect. For Mercury, the gravitoelectric precession of its perihelion is 43.98 arcseconds per century; the gravitomagnetic one is about $-3$ milli-arcseconds per century (mas cy$^{-1}$), which is comparable with its $\dot{\omega}_{\mathrm{sup}}$ (see Table \ref{Tab:omegasup}). In the classical test, $\delta^{\mathrm{GR}}_{\dot{\omega}}$ of Eq. (\ref{dPS}) only includes the gravitoelectric perihelion shift caused by $M_{\odot}$. Hence, we add the Lense-Thirring term $\dot{\omega}_{\,\mathrm{LT}}$ to Eq. (\ref{domegasupp}) and it is \cite{Lense1918PhyZ19.156}
\begin{equation}
  \label{domegaLT}
  \dot{\omega}_{\,\mathrm{LT}} = -\frac{6GS_{\odot}\cos i_{\mathrm{P}}}{c^2a_{\mathrm{P}}^3(1-e_{\mathrm{P}}^2)^{3/2}},
\end{equation}
where $ S_{\odot} = 1.9 \times 10^{41}$ kg m${}^2$ s${}^{-1}$ \cite{Pijpers2003AA402.683} and $i_{\mathrm{P}}$ is the inclination of the planetary orbit to the equator of the Sun. The uncertainty of $S_{\odot}$ is currently about $1\%$ \cite{Pijpers2003AA402.683}. This effect of the Sun on the planetary motions has been studied in several works \cite{Iorio2005AA433.385,Iorio2011ApSS331.351,Iorio2012SoPh281.815}. Equation (\ref{domegaLT}) only holds in a coordinate system whose $z$ axis is aligned with the Sun's angular momentum. A general formula for an arbitrary orientation can be found in Refs. \cite{Iorio2011PRD84.124001,Iorio2012GRG44.719}. It is useful in extrasolar planets and black holes, for which the orientation of the spin axis is generally unknown. We add the third term in Eq. (\ref{domegasupp}) to include the dimensionless uncertainty of the Sun's quadrupole moment $\mathcal{J}_{\odot}$ \cite{Iorio2005AA431.385}, which is currently about $\pm 10\%$ \cite{Damiani2011JASTP73.241,Pireaux2003ApSS284.1159,Rozelot2004astro.ph0403382,Rozelot2011EPJH36.407,Rozelot2013SoPh287.161}. The Sun's quadrupole moment in INPOP10a is fitted to observations as $ J^{\odot}_2 = ( 2.40 \pm 0.25 ) \times 10^{-7}$ \cite{Fienga2011CMDA111.363} and its value in EPM2011 is $ J^{\odot}_2 = (2.0 \pm 0.2) \times 10^{-7}$ \cite{Pitjeva2013SSR47.386}. This uncertainty of $J^{\odot}_2$ can cause an extra precession for a planet, which is \cite{Kozai1959AJ64.367}
\begin{equation}
  \label{}
  \dot{\omega}_{\mathcal{J}_{\odot}} = \frac{3}{2}\mathcal{J}_{\odot}\frac{J^{\odot}_2 R_{\odot}^2}{a_{\mathrm{P}}^2(1-e_{\mathrm{P}}^2)^2}n_{\mathrm{P}}\bigg(2-\frac{5}{2}\sin^2i_{\mathrm{P}}\bigg),
\end{equation}
where $R_{\odot}$ is the Sun's radius.  It is clearly showed \cite{Liang2014RAA14.527,Deng2015NewAst35.36} that although the uncertainty of $J^{\odot}_2$ can barely affect the outer planets, such as Jupiter and Saturn, but it will significantly change the dynamics of the inner planets, especially Mercury. The higher order multipoles like $J^{\odot}_3$ and $J^{\odot}_4$ have negligible impacts on the perihelion precessions \cite[e.g.][]{Renzetti2013JAA34.341,Renzetti2014Ap&SS352.493}.

INPOP10a \citep{Fienga2011CMDA111.363} ephemeris provides $\dot{\omega}_{\mathrm{sup}}$ for some planets in the Solar System: Mercury, Venus, Earth-Moon Barycenter (EMB), Mars, Jupiter and Saturn. Similarly, EPM2011 \citep{Pitjeva2013SSR47.386} also gives those values of the planets from Mercury to Saturn. These numbers are taken from Ref. \cite{Fienga2011CMDA111.363} and Refs. \cite{Pitjeva2013MNRAS432.3431, Pitjev2013AstL39.141} respectively (see Table \ref{Tab:omegasup} for details). It can be found that $\dot{\omega}_{\mathrm{sup}}$ of Mercury and Venus from EPM2011 are considerably larger than those of INPOP10a, while Venus and Jupiter have non-zero values of $\dot{\omega}_{\mathrm{sup}}$ in EPM2011.

By using the method of weighted least squares, we simultaneously estimate the bounds on $b$ and $\mathcal{J}_{\odot}$ with \textit{all} the planets in Table \ref{Tab:omegasup}. We find that (i) INPOP10a yields the bounds as $b = (-0.8 \pm 7.6 )\times 10^{-4}$  and $\mathcal{J}_{\odot} = (6.5\pm9.1)\%$; and (ii) EPM2011 gives $b= (1.9 \pm 2.5 )\times 10^{-4}$ and $\mathcal{J}_{\odot} = (2.0\pm 8.4)\%$. Our results are at least 100 times tighter than previous results \cite{Liu2000ApJ538.386} in which Mercury's perihelion precession was only considered. These results are summarized in Table \ref{Tab:summaryresults}. The results obtained by INPOP10a and EPM2011 are compatible with each other. Furthermore, the values of $\mathcal{J}_{\odot}$ given by INPOP10a and EPM2011 are compatible with the current uncertainty of $\pm 10\%$.

\begin{table}
\begin{center}
\caption[]{Supplementary advances in the perihelia $\dot{\omega}_{\mathrm{sup}}$ given by INPOP10a and EPM2011.}
\label{Tab:omegasup}
\begin{tabular}{lcc}
  \hline\noalign{\smallskip}
 & \multicolumn{2}{c}{$\dot{\omega}_{\mathrm{sup}}$ (mas cy${}^{-1}$)}  \\
 \cline{2-3} \noalign{\smallskip}
  & INPOP10a ${}^a$ & EPM2011 ${}^b$ \\
  \hline\noalign{\smallskip}
  Mercury & $0.4\pm0.6$  &  $-2.0 \pm 3.0$\\
  Venus & $0.2\pm1.5$  & $2.6 \pm 1.6$\\
  EMB & $-0.2\pm0.9$ & --\\
  Earth & -- & $0.19 \pm 0.19$\\
  Mars & $-0.04\pm0.15$ & $-0.020 \pm 0.037$\\
  Jupiter & $-41\pm42$ & $58.7 \pm 28.3$\\
  Saturn & $0.15\pm0.65$ & $-0.32 \pm 0.47$ \\
  \noalign{\smallskip}\hline
\end{tabular}
\end{center}
Notes: ${}^a$Taken from Ref. \cite{Fienga2011CMDA111.363}. ${}^b$Provided by Refs. \cite{Pitjeva2013MNRAS432.3431,Pitjev2013AstL39.141}.
\end{table}

\subsection{Light deflection}

In astrometric observation for gravitational light bending, the Very Long Baseline Array (VLBA) demonstrated the accuracy of measuring relative positions of radio sources can reach $\sim 10$ to 100 $\mu$as \citep{Fomalont2003ApJ598.704,Fomalont2009ApJ699.1395}, which makes us have, from Eq. (\ref{dLD}),
\begin{equation}
  \label{qdGRphi}
   1.75 \times 10^{5} \bigg|\sqrt{1-\frac{3b^{2}}{4}}-1\bigg| \bigg(\frac{d}{R_{\odot}}\bigg)^{-1} \le \frac{\delta^{\mathrm{GR}}_{\Delta \phi}}{10\; \mu\mathrm{as}}.
\end{equation}
By taking $d\sim5R_{\odot}$ and $\delta^{\mathrm{GR}}_{\Delta t_{\mathrm{IC}}}\sim 10\mu as$, we can obtain a new upper bounds on $b$ from light deflection as $|b|<8.7\times10^{-3}$, which is at least 10 times tighter than the previous result of Ref. \cite{Liu2000ApJ538.386} (see Table \ref{Tab:summaryresults} for a summary).

\subsection{Gravitational time delay at IC}

We assume the receiver is carried by a Venus's spacecraft and the emitter is on the Earth. The uncertainty of the receiver's position is about several centimeters; and the uncertainty of emitter's position is at the level of several hundreds meters. Their contribution is about $1$ microsecond ($\mu$s) according to Eq. (\ref{tt}). It can impose a bound on $b$ as
\begin{equation}
  \label{qdGRtIC}
 19.7\bigg| \sqrt{1-\frac{3b^{2}}{4}}-1 \bigg|\ln\frac{r_{E}}{r_{R}} \le  \frac{\delta^{\mathrm{GR}}_{\Delta t_{\mathrm{IC}}}}{1\;\mu\mathrm{s}}.
\end{equation}
By taking $r_{E}\sim1$ au, $r_{R}/r_{E}\sim0.7$ and $\delta^{\mathrm{GR}}_{\Delta t_{\mathrm{IC}}}\sim 1$ $\mu$s, we obtain $|b|<0.61$. No useful bound can be placed on $b$ from the time delay at IC.

\subsection{\textit{Cassini} SC experiment}

In the \textit{Cassini} SC experiment \citep{Bertotti2003Nature425.374}, the deviation from GR is $\delta^{\mathrm{GR}}_{y_{\mathrm{SC}}} \le 10^{-14}$. This result was achieved by using a multi-frequency link. From Eq. (\ref{dySC}), we have
\begin{equation}
  \label{qdGRySC}
   1.01\times10^{5}\bigg|\sqrt{1-\frac{3b^{2}}{4}}-1 \bigg| \le  \frac{\delta^{\mathrm{GR}}_{y_{\mathrm{SC}}}}{10^{-14}}.
\end{equation}
By taking $\delta^{\mathrm{GR}}_{y_{\mathrm{SC}}}\sim 10^{-14}$, we can obtain an upper bound on $b$ from frequency shift as $|b|<5.1\times10^{-3}$, which is at least 10 times stronger than the one of Ref. \cite{Liu2000ApJ538.386}.

\begin{table}
\begin{center}
	\caption[]{Summary of $b$ estimated by various experiments and observations.}
\label{Tab:summaryresults}
\begin{tabular}{lrll}
  \hline\noalign{\smallskip}
  Experiment/Observation &  $b$ $(10^{-4})$ & Datasets & Ref. \\
  \hline\noalign{\smallskip}
  Perihelion shfit & $-300\pm700$ & Mercury \cite{Shapiro1976PRL36.555}& \cite{Liu2000ApJ538.386}\\
  & $-0.8 \pm 7.6 $ & INPOP10a${}^a$ \cite{Fienga2011CMDA111.363} & This work  \\
  & $ 1.9 \pm 2.5 $ &  EPM2011${}^a$ \cite{Pitjeva2013MNRAS432.3431} & This work \\
  Light deflection & 700 & Sun \cite{Robertson1991Natur349.768} & \cite{Liu2000ApJ538.386} \\
   & 87 & Sun \cite{Fomalont2003ApJ598.704,Fomalont2009ApJ699.1395} & This work \\
  Time delay at SC & 700 & Mars \cite{Reasenberg1979ApJ234.219} & \cite{Liu2000ApJ538.386} \\
  Time delay at IC & 6100 &  Venus & This work \\
  \textit{Cassini} SC & 51 & {\it Cassini} \cite{Bertotti2003Nature425.374} & This work \\
  Geodetic precession & 200 & GP-B \cite{Everitt2011PRL106.221101} & \cite{Overduin2013GRG45.1723} \\
  \noalign{\smallskip}\hline
\end{tabular}
\end{center}
Notes: ${}^a$The results are obtained according to all the planets in Table \ref{Tab:omegasup}.
\end{table}

\section{Conclusions and discussion}

\label{sec:con}

In this work, as an extension of previous works on classical tests of Kaluza-Klein (KK) gravity and as an attempt to find more stringent constraints on this theory, we investigate its effects on physical experiments and astronomical observations conducted in the Solar System by modeling new observable effects, using improved models for confronting theoretical prediction with observations and adopting new datasets.

First, we calculate gravitational time delay at IC caused by KK gravity, which was not considered in the previous work of Ref. \cite{Liu2000ApJ538.386}. The time delay measurements at SC are highly dominated by the noise due to solar corona, but the biggest uncertainties at IC come from the positions of the receiver and the emitter. Second, the \textit{Cassini} SC experiment \cite{Bertotti2003Nature425.374} is, for the first time, used to test KK gravity. We calculate the fractional frequency shift caused by KK gravity for a two-way radio signal. Third, compared to previous works, we refine the model, which confronts the perihelion shift induced by KK gravity with modern Solar System ephemerides INPOP10a and EPM2011, by taking the Lense-Thirring effect due to the Sun's angular momentum and the uncertainty of the Sun's quadrupole moment into account. These two factors were absent previously.

With these efforts and with new datasets, we find improved and new upper bounds on the model parameter $b$ of KK gravity (see Tab. \ref{Tab:summaryresults} for a summary), although it is shown that time delay experiments at IC is not quite suitable for testing it with the currently limited accuracy of ranging measurements. With new observation of light deflection by the Sun, we obtain $|b|<8.6\times10^{-3}$. The \textit{Cassini} SC experiment gives a upper bound as $|b|<5.1\times10^{-3}$. Based on the supplementary advances of the perihelia provided by INPOP10a and EPM2011 ephemerides, we obtain our best upper bounds on $b$: $b=(-0.8\pm7.6)\times10^{-4}$ from INPOP10a and $b=(1.9\pm2.5)\times10^{-4}$ from EPM2011. Both of them are tighter than the one of Ref. \cite{Liu2000ApJ538.386} by at least 2 orders of magnitude. In order to achieve these bounds, we take the Lense-Thirring effect due to the Sun's angular momentum and the uncertainty of the Sun's quadrupole moment into account. These two factors were not considered in previous works.

With these new upper bounds we obtained that $|b|\lesssim10^{-4}$, it means that the fifth-dimensional space in the soliton case is a very flat extra dimension in the Solar System, even in the vicinity of the Sun where the light rays pass through in the light deflection observations and the \textit{Cassini} SC experiment. A very small $b$ also suggests the soliton metric is very close to four-dimensional Schwarzschild metric plus a flat and therefore physically innocuous fifth dimension. Our results show KK gravity still remains consistent with current Solar System experiments and observations, but has a much smaller room to survive. A question is often raised whether these and future more and more tighter bounds can conclusively rule out KK gravity. We do not think so because these tests are based on the soliton solution, whose physical nature remains controversial (see Ref. \cite{Wesson2011arXiv1104.3244} for a review). As pointed out by the authors of Ref. \cite{Overduin2013GRG45.1723}, even the null result should be taken care of with an open mind and they can highlight the need for new solutions in five dimensions and for a generalization of Birkhoff's theorem.

Several open issues remains in testing KK gravity. One is to test its effects in the vicinity of the Earth. In the future, it may be possible by tracking a drag-free satellite with laser ranging and time transfer links \cite{Damour1994PRD50.2381,Deng2013MNRAS431.3236}.

\begin{acknowledgements}
The work of XMD is supported by the National Natural Science Foundation of China under Grant Nos. 11473072 and 11533004, the Fundamental Research Program of Jiangsu Province of China under Grant No. BK20131461. The work of YX is funded by the Natural Science Foundation of China under Grant No. 11573015, the Opening Project of Shanghai Key Laboratory of Space Navigation and Position Techniques under Grant No. 14DZ2276100.
\end{acknowledgements}

\bibliographystyle{spphys}       
\bibliography{Refs20151010}   

%
%

\end{document}